\documentclass[aps,prb,twocolumn,superscriptaddress,groupedaddress,amsmath,amssymb]{revtex4-2}
\usepackage[dvipdfmx]{graphicx} 
\usepackage{ulem}

\usepackage{graphicx, color}
\usepackage{float}
\usepackage{dcolumn}
\usepackage{bm}
\usepackage{color}

\begin{document}

\title{Valley-dependent electron-phonon scattering in thermoelectric semimetal Ta$_2$PdSe$_6$}
\author{
Masayuki Ochi$^{1,2}$,
Hitoshi Mori$^3$, 
and Akitoshi Nakano$^4$ 
}
\affiliation{
$^1$Department of Physics, The University of Osaka, Toyonaka, Osaka 560-0043, Japan\\
$^2$Forefront Research Center, The University of Osaka, Toyonaka, Osaka 560-0043, Japan\\
$^3$Institute for Materials Research, Tohoku University, Katahira, Aoba-ku, Sendai 980-8577, Japan\\
$^4$Department of Applied Physics, Graduate School of Engineering, Nagoya University, Furo-cho, Chikusa-ku, Nagoya, Aichi 464-8603, Japan
}
\date{\today}
\begin{abstract}
Quasi-one-dimensional transition-metal chalcogenide
Ta$_2$PdSe$_6$ is a promising thermoelectric semimetal due to the strong electron-hole asymmetry in the carrier lifetime. However, the microscopic origin of such a strong asymmetry remains unclear.
In this study, we theoretically investigate electron-phonon scattering in Ta$_2$PdSe$_6$.
There is a soft phonon mode mainly consisting of atomic displacements in PdSe$_4$ chains.
This soft mode is strongly coupled with the highest valence band at the $\Gamma$ point, which lies slightly below the Fermi energy, and causes strong electron-phonon scattering.
The bottom of the electron pocket energetically overlapped with that band also suffers from strong intervalley scattering, by which the imaginary part of the electron self-energy exhibits a sharp change near the Fermi level.
On the other hand, the imaginary part of the self-energy for carriers in the hole pocket shows a moderate energy dependence.
Thus, we find that electron-phonon scattering is strongly valley-dependent.
Our finding will help us to understand the distinctive transport properties observed in Ta$_2$PdSe$_6$.
\end{abstract}

\maketitle

\section{Introduction}

High-performance thermoelectric materials have long been looked for mainly in slightly doped insulators or semiconductors.
The reason is that electron and hole carriers (i.e., carriers above or below the chemical potential, respectively) contribute oppositely to the Seebeck effect, by which metallic systems tend to have a small Seebeck coefficient.
However, the above consideration leads to the idea that metallic systems with strong electron-hole asymmetry can offer an unexplored arena for efficient thermoelectric conversion~\cite{Markov2019}.
In particular, it has received strong attention that electron-hole asymmetry in relaxation time can significantly change transport properties and sometimes results in a high thermoeletric power factor and a high dimensionless figure of merit $ZT$.
In Ref.~[\onlinecite{Xu2014_Li}], the positive sign of the Seebeck coefficient in Li, which was not reproduced by the constant relaxation time approximation, was explained in terms of the energy-dependent relaxation time due to the van Hove singularity slightly above the Fermi energy.
Here, the DOS peak slightly above the Fermi energy induces a strong scattering of electron carriers, and thus the Seebeck coefficient becomes positive.
In Ref.~[\onlinecite{Xu2014}], it was proposed that topological insulators with surface bands can have a high $ZT$ when the chemical potential lies just below the bulk bands due to the strong asymmetry in the relaxation time of topological surface states between long-lived hole carriers that do not have energetic overlap with the bulk bands and electron carriers that suffer from strong scattering with the bulk bands.
As proposed in this case, energetic overlap with heavy bands can energy-selectively shorten the carrier lifetime and thus enhance the Seebeck coefficient, similarly to energy filtering~\cite{Heremans2005,Zeng2007,Faleev2008}.
This view is closely related to the concept of the boxcar-shaped transport function that optimizes $ZT$ as investigated by a seminal work by Mahan and Sofo~\cite{Mahan1996} and the following studies of it~\cite{Whitney2014,Maassen2021,Park2021_1,Park2022,Ding2023}.

Based on this strategy, electron-hole-asymmetric relaxation time has been theoretically found in several systems such as CoSi and a monolayer MoS$_2$ allotrope, both of which have Dirac bands with a heavy parabolic band above or below the chemical potential~\cite{Xia2019_CoSi,Xia2019}, and
multi-valley half-Heusler alloys with strong interband scattering~\cite{Kumarasinghe2019,Fedorova2022}.
A two-band model with band offset, where one band plays the role of the scatterer and the other offers carriers with electron-hole-asymmetric relaxation time due to interband scattering, was investigated from this perspective~\cite{Kumarasinghe2019,Fedorova2022,Ochi2023,Graziosi2024}.
The concept of the multi-valley band structure that has been regarded as a favorable feature of thermoelectric materials~\cite{Pei2011} has now been partially reconsidered, since band convergence shortens the carrier lifetime~\cite{Kumarasinghe2019,Ponce2019,Ponce2019_PRL,Sohier2019,Park2021}.
It was theoretically pointed out that bipolar conduction in narrow-gap semiconductors with strong electron-hole asymmetry can lead to a high thermoelectric power factor~\cite{Graziosi2020,Graziosi2022}.
Recently, high thermoelectric performance via interband scattering was experimentally reported in NiAu alloys~\cite{Garmroudi2023, Riss2024}.
It was also found that a large Seebeck coefficient is realized in Ni$_3$In~\cite{Garmroudi2025} and Ni$_3$Ge~\cite{Garmroudi2025_Ge} due to the Kagome-lattice configuration of Ni atoms, by which carriers in dispersive bands are energy-selectively scattered by localized Ni-$d$ bands and thus exhibit an electron-hole-asymmetric lifetime.
It was theoretically pointed out that antisite defects in Fe$_2$VAl can induce energy-selective scattering and a resulting sign change of the Seebeck coefficient~\cite{Tohyama2025}.

The quasi-one-dimensional transition-metal chalcogenide
Ta$_2$PdSe$_6$~\cite{Keszler1985} is a promising thermoelectric semimetal that exhibits high thermoelectric performance at low temperatures~\cite{Nakano2021,Nakano2021_JPSJ,Nakano2022}.
Ta$_2$PdSe$_6$ has the Peltier conductivity of 100 A cm$^{-1}$K$^{-1}$ at 10 K and the thermoelectric power factor of 2.4 mW cm$^{-1}$K$^{-2}$ at 15 K~\cite{Nakano2021}, both of which are exceptionally large.
It was experimentally found that the mobilities of electron and hole carriers differ by an order of magnitude at low temperature~\cite{Nakano2022,Nakano2025}.
A recent angle-resolved photoemission spectroscopy (ARPES) study~\cite{Ootsuki2025} found a replica feature of the electron band, which suggests a strong electron-boson interaction and also electron-hole asymmetry, since this feature is only observed for the electron pocket. In addition, a kink of the band structure possibly due to electron-phonon coupling was also observed.
For theoretical studies of Ta$_2$PdSe$_6$, while the electronic band structure and the phonon band structure have been investigated in some studies~\cite{Nakano2021,Yang2022,Kato2024},
the scattering properties of the carriers have not been investigated much.

In this study, we theoretically investigate electron-phonon scattering in Ta$_2$PdSe$_6$.
There is a soft phonon mode consisting of atomic displacements in PdSe$_4$ chains.
This soft mode is strongly coupled with the highest valence band at the $\Gamma$ point, which lies slightly below the Fermi energy, and causes strong electron-phonon scattering.
The bottom of the electron pocket also suffers from strong intervalley scattering, by which the imaginary part of the electron self-energy exhibits a sharp change near the Fermi level.
On the other hand, the imaginary part of the self-energy for carriers in the hole pocket shows a moderate energy dependence.
We find that electron-phonon scattering is strongly valley-dependent.
Although our calculation results do not fully explain the observed electron-hole asymmetry in experiments, our finding will offer important knowledge to understand the distinctive transport properties of Ta$_2$PdSe$_6$.

This paper is organized as follows.
Section~\ref{sec:methods} describes the theoretical methods used in this study. Our calculation results are shown in Sec.~\ref{sec:results}.
First, we present first-principles electronic and phonon band structures in Sec.~\ref{sec:bands}. Next, electron self-energy by electron-phonon scattering is analyzed in Sec.~\ref{sec:self}. 
Some discussion of the relationship between our calculation results and experimentally observed transport properties is presented in Sec.~\ref{sec:discussion}.
Section~\ref{sec:summary} summarizes this study.

\section{Methods\label{sec:methods}}

First-principles calculations based on density functional theory were performed using the Perdew-Burke-Ernzerhof parametrization of the generalized gradient approximation~\cite{PBE} and the projector-augmented wave (PAW)~\cite{PAW} method as implemented in Quantum ESPRESSO~\cite{QE2009,QE2017,QE2020}.
We used PAW pseudopotentials provided in pslibrary~\cite{pslibrary, pslibrary_web}.
The plane-wave cutoff energies for Kohn--Sham orbitals and charge density are 50 and 500 Ry, respectively. Gaussian smearing with a smearing width of 0.01 Ry was applied. Spin-orbit coupling was not included in our calculations because of the expensive computational cost for electron-phonon coupling.
We verified that the electronic band dispersion is not largely affected by including spin-orbit coupling.

\begin{figure}
  \begin{center}
  \includegraphics[width = \linewidth]{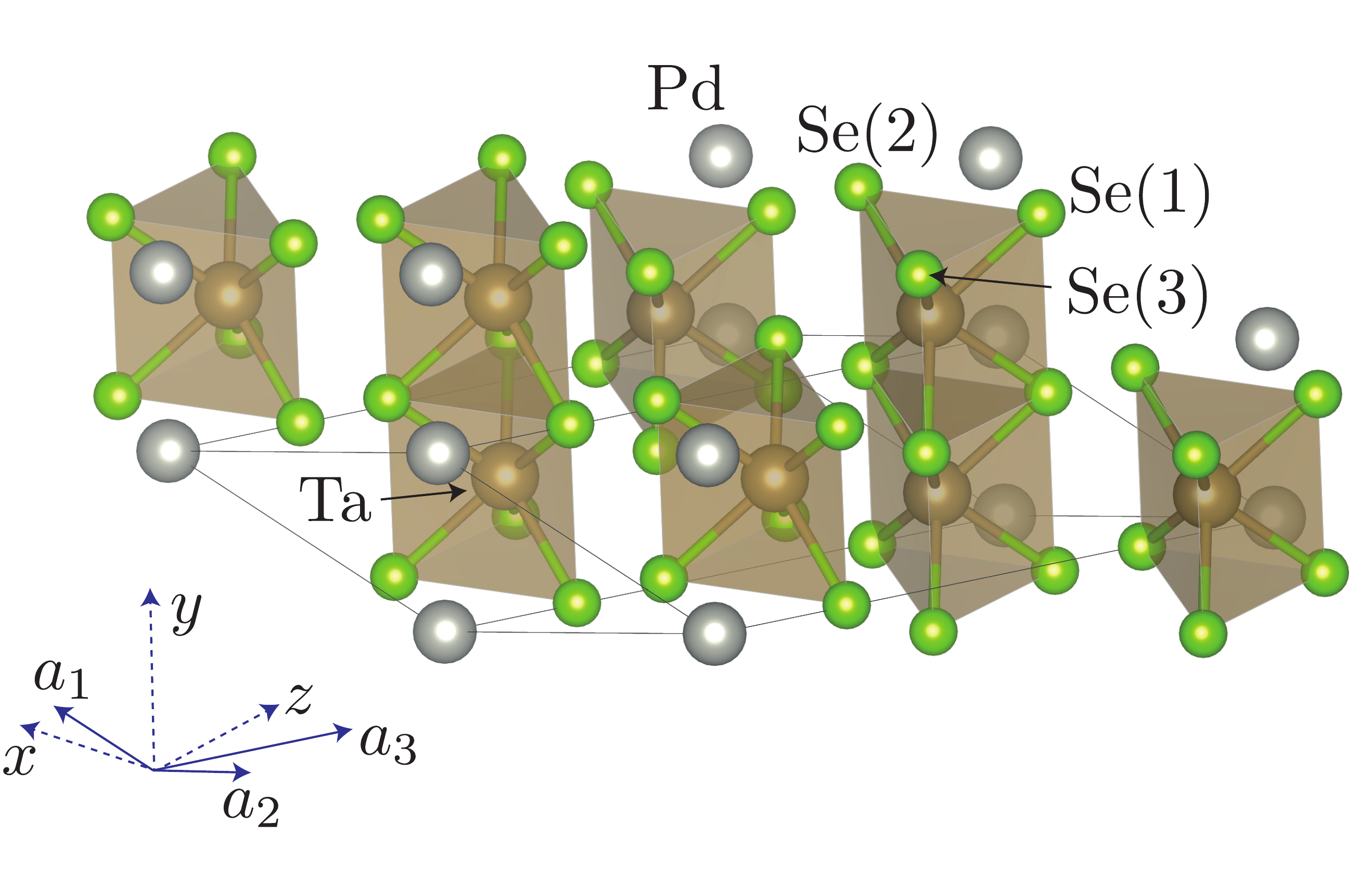}
  \caption{
  Crystal structure of Ta$_2$PdSe$_6$ depicted using the VESTA software~\cite{vesta}.
  }
  \label{fig:crystal}
  \end{center}
\end{figure}

First, we optimized the atomic coordinates while using experimental lattice constants ($a=12.4179$ \AA, $b=3.3691$ \AA, $c=10.3951$ \AA, and $\beta=116.192^\circ$ with the $C2/m$ space group) taken from Ref.~[\onlinecite{Nakano2021}].
Hellmann--Feynman forces acting on the atoms were less than $5\times 10^{-4}$ eV/\AA\ for the optimized crystal structure shown in Fig.~\ref{fig:crystal}.
The lattice vectors in Cartesian coordinates are
\begin{equation}
{\bm a}_1 =
\begin{pmatrix} a/2 \\ b/2 \\ 0 \end{pmatrix},\ \ 
{\bm a}_2 =
\begin{pmatrix} -a/2 \\ b/2 \\ 0 \end{pmatrix},\ \ 
{\bm a}_3 =
\begin{pmatrix} c \cos \beta \\ 0 \\ c\sin \beta \end{pmatrix}.
\end{equation}
As shown in Fig.~\ref{fig:crystal}, Ta$_2$PdSe$_6$ has one-dimensional chains along the $y$ direction. The reciprocal lattice vectors in Cartesian coordinates are
\begin{gather}
{\bm b}_1 =
\begin{pmatrix} 2\pi/a \\ 2\pi/b \\ -2\pi/(a\tan \beta) \end{pmatrix},\ \ 
{\bm b}_2 =
\begin{pmatrix} -2\pi/a \\ 2\pi/b \\ 2\pi/(a\tan \beta) \end{pmatrix},\notag \\
{\bm b}_3 =
\begin{pmatrix} 0 \\ 0 \\ 2\pi /(c \sin \beta) \end{pmatrix}.
\end{gather}
Electronic structure calculations were performed using a $10\times 10\times 4$ ${\bm k}$ mesh.
We performed a phonon calculation based on density functional perturbation theory using a $10\times 10\times 4$ ${\bm q}$ mesh.
 
After these calculations, we calculated the electron-phonon coupling with Wannier interpolation as implemented in EPW~\cite{EPW}. For Wannierization of electronic wave functions, we constructed Wannier orbitals for the Ta-$d$, Pd-$d$, and Se-$p$ orbitals using Wannier90~\cite{Pizzi2020}. Outer and inner energy windows were set as [$-7$:7] and [$-7$:$1.8$] eV, respectively, where the Fermi energy was set to zero.
After Wannierization, the electron self-energy originating from the electron-phonon coupling was evaluated with a $160\times 160\times 16$ ${\bm q}$-mesh for some sets of ${\bm k}$ points as described later, at temperature $T = 300$ K. The electron self-energy $\Sigma_{n{\bm k}}$ for the Kohn--Sham state at the $n$-th band and wavevector ${\bm k}$ was calculated as
\begin{gather}
\Sigma_{n{\bm k}} = \sum_{{\bm q}\nu,m} |g_{mn,\nu}({\bm k}, {\bm q})|^2 \label{eq:self} \\
\times
\bigg[ \frac{n(\omega_{{\bm q}\nu})+f(\epsilon_{m{\bm k}+{\bm q}})}{\epsilon_{n{\bm k}}- \epsilon_{m{\bm k}+{\bm q}} + \omega_{{\bm q}\nu} -i\eta} +
\frac{n(\omega_{{\bm q}\nu})+ 1 -f(\epsilon_{m{\bm k}+{\bm q}})}{\epsilon_{n{\bm k}}- \epsilon_{m{\bm k}+{\bm q}} - \omega_{{\bm q}\nu} -i\eta}
\bigg] \notag
\end{gather}
where $g_{mn,\nu}({\bm k}, {\bm q})$, $n$, $f$, $\epsilon$, and $\omega$ are the electron-phonon matrix element that describes the scattering from the electronic state ($n$, ${\bm k}$) to the electronic state ($m$, ${\bm k}+{\bm q}$) 
mediated by the phonon in branch $\nu$ and wavevector ${\bm q}$, Bose--Einstein distribution function, Fermi--Dirac distribution function, Kohn--Sham energy, and phonon energy, respectively.
The small positive parameter $\eta$ in Eq.~(\ref{eq:self}) called degaussw in EPW was set as 0.015 eV, for which we verified that using different values, 0.01 and 0.02 eV, makes a negligible difference. The Fermi level at $T=300$ K was 2.6 meV above the Fermi energy at zero temperature, both of which were evaluated using the tight-binding model consisting of the Wannier orbitals with a $420\times 420\times 16$ ${\bm k}$-mesh.

\section{Results\label{sec:results}}

\subsection{Electronic and phonon band structures\label{sec:bands}}

\begin{figure}
  \begin{center}
  \includegraphics[width = 0.9\linewidth]{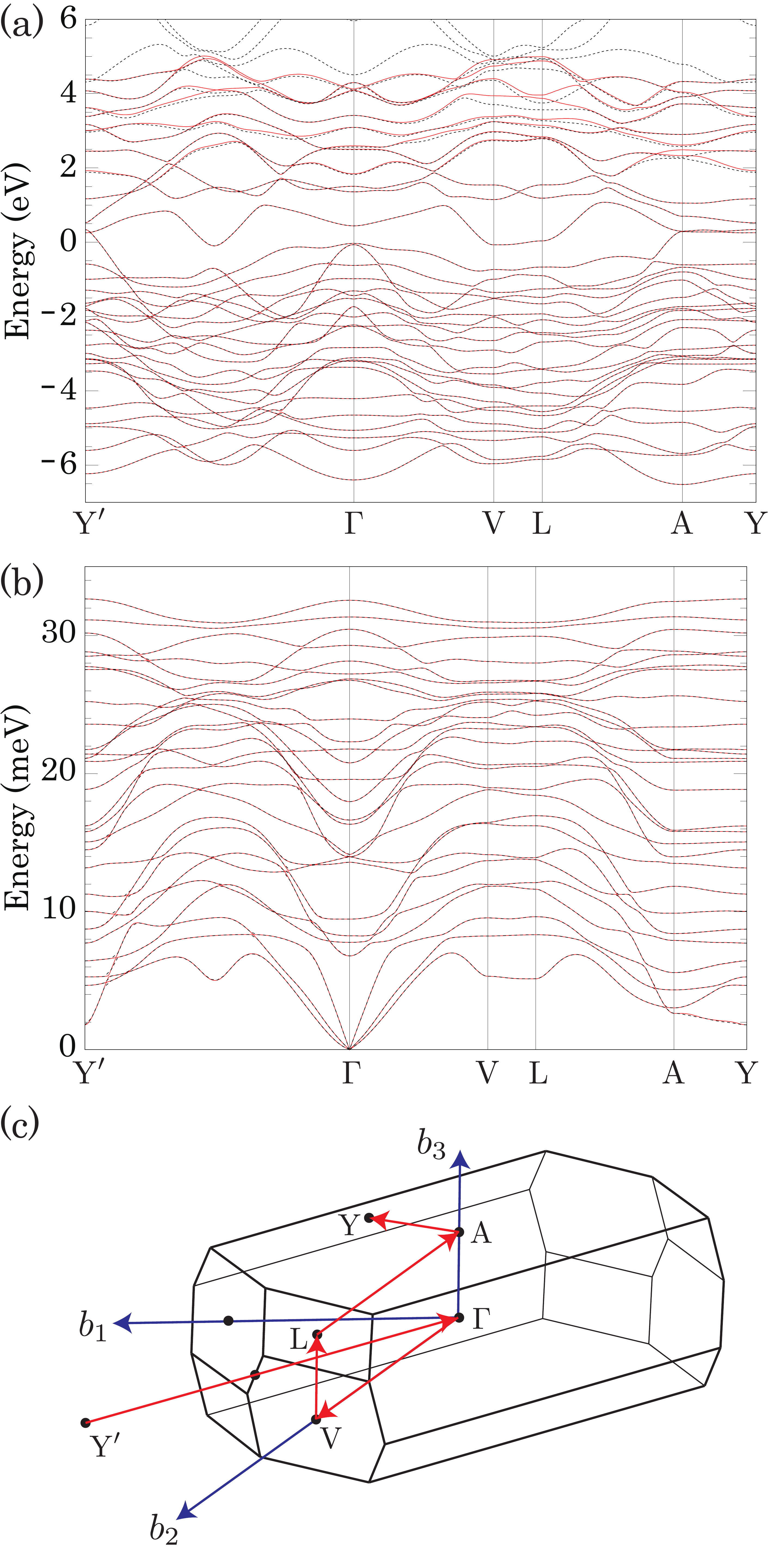}
  \caption{
  (a) Electronic band structure and (b) phonon band structure along the ${\bm k}$ path shown in the Brillouin zone (c).
  In panels (a)--(b), black broken and red solid lines represent first-principles band structure and that obtained by Wannier interpolation, respectively. Definitions of special ${\bm k}$ points are as follows: Y$'=0.5({\bm b}_1 + {\bm b}_2)$, V$=0.5 {\bm b}_2$, L$=0.5({\bm b}_2 + {\bm b}_3)$, A$=0.5 {\bm b}_3$, and Y$=0.5({\bm b}_1 - {\bm b}_2)$. Note that Y and Y$'$ are equivalent while the latter is not in the first Brillouin zone.
  }
  \label{fig:band}
  \end{center}
\end{figure}

\begin{figure}
  \begin{center}
  \includegraphics[width = 0.9\linewidth]{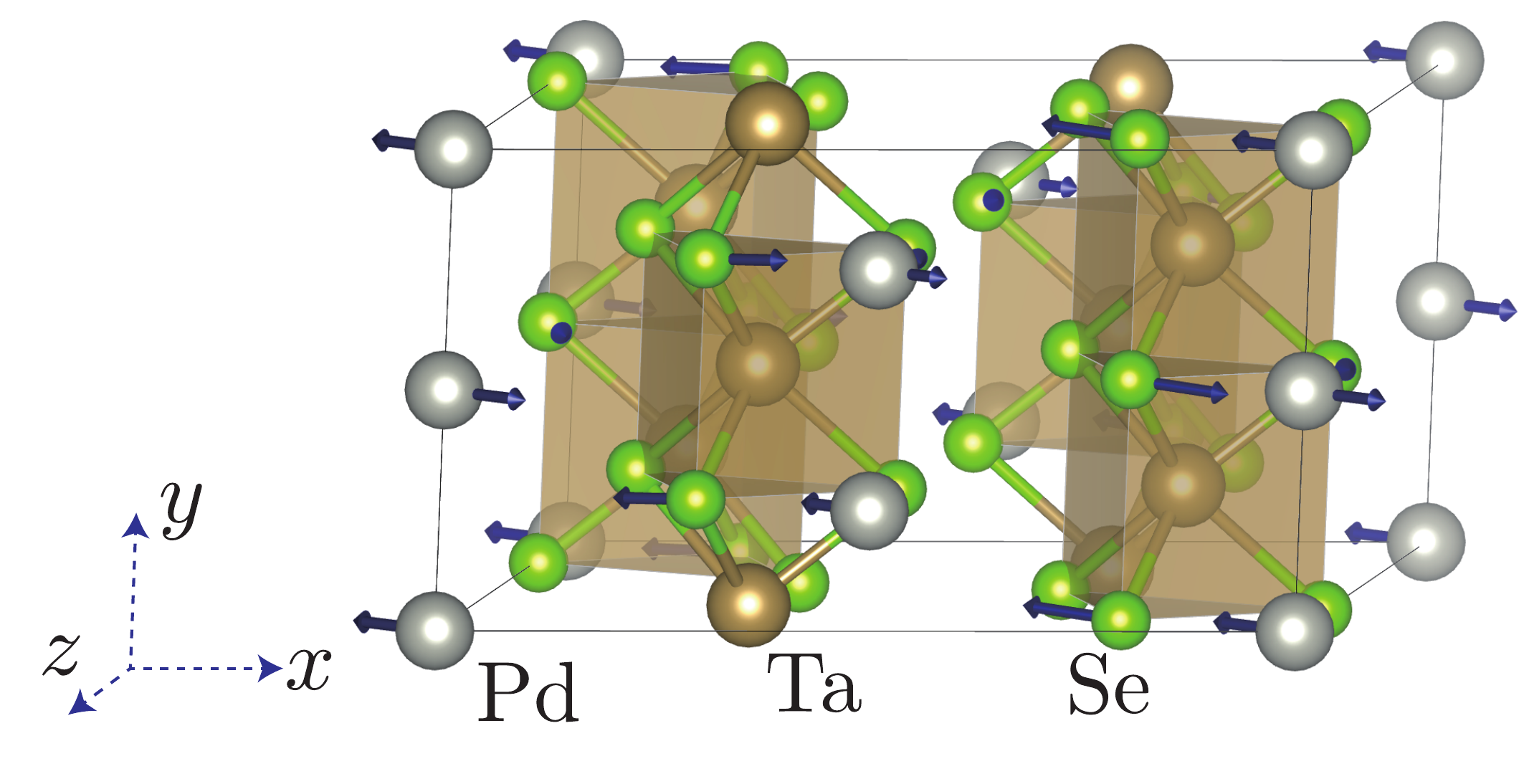}
  \caption{
   The lowest (soft) phonon mode at the midpoint of the Y$'$-$\Gamma$ line, i.e., ${\bm q}=0.25({\bm b}_1+{\bm b}_2)$.
  }
  \label{fig:soft_mode}
  \end{center}
\end{figure}

First-principles electronic and phonon band structures are shown in Figs.~\ref{fig:band}(a) and (b), respectively, with black broken lines. The ${\bm k}$ or ${\bm q}$ path taken here is shown in Fig.~\ref{fig:band}(c).
As discussed in previous studies~\cite{Nakano2021,Nakano2021_JPSJ,Nakano2022,Ootsuki2025,Yang2022,Kato2024}, the electronic band structure is semimetallic.
The phonon band structure has no imaginary mode, by which we confirmed that the experimentally reported space group is consistently stable in our calculation.
On the other hand, we found a slight softening around the midpoint of the Y$'$-$\Gamma$ line with a phonon energy of $\sim 5$ meV.
The same softening is also found along the V-L line, which also has $q_y\sim \pi/b$, due to the quasi-one-dimensionality of the system.
Figure~\ref{fig:soft_mode} shows the atomic displacement vectors of the soft phonon mode at the midpoint of the Y$'$-$\Gamma$ line, i.e., ${\bm q}=0.25({\bm b}_1+{\bm b}_2) = (0, \pi/b, 0)$ in Cartesian coordinates, visualized using
Re[$C {\bm u} \mathrm{e}^{i{\bm q}\cdot {\bm R}}$], where ${\bm u}$ is the phonon eigenvector, ${\bm R}$ is the unit-cell position, and $C$ is a complex constant chosen for visualization.
This soft phonon mode is transverse and its atomic components are mainly Pd and surrounding Se, i.e., Se(1) and Se(2).
This mode has an eigenvalue of $-1$ with respect to the $C_2$ rotational symmetry in the $xz$ plane, i.e., $(x,y,z) \to (-x, y,-z)$.
We shall discuss the relation between this soft mode and the electronic band structure later in this paper.

\begin{figure}
  \begin{center}
  \includegraphics[width = \linewidth]{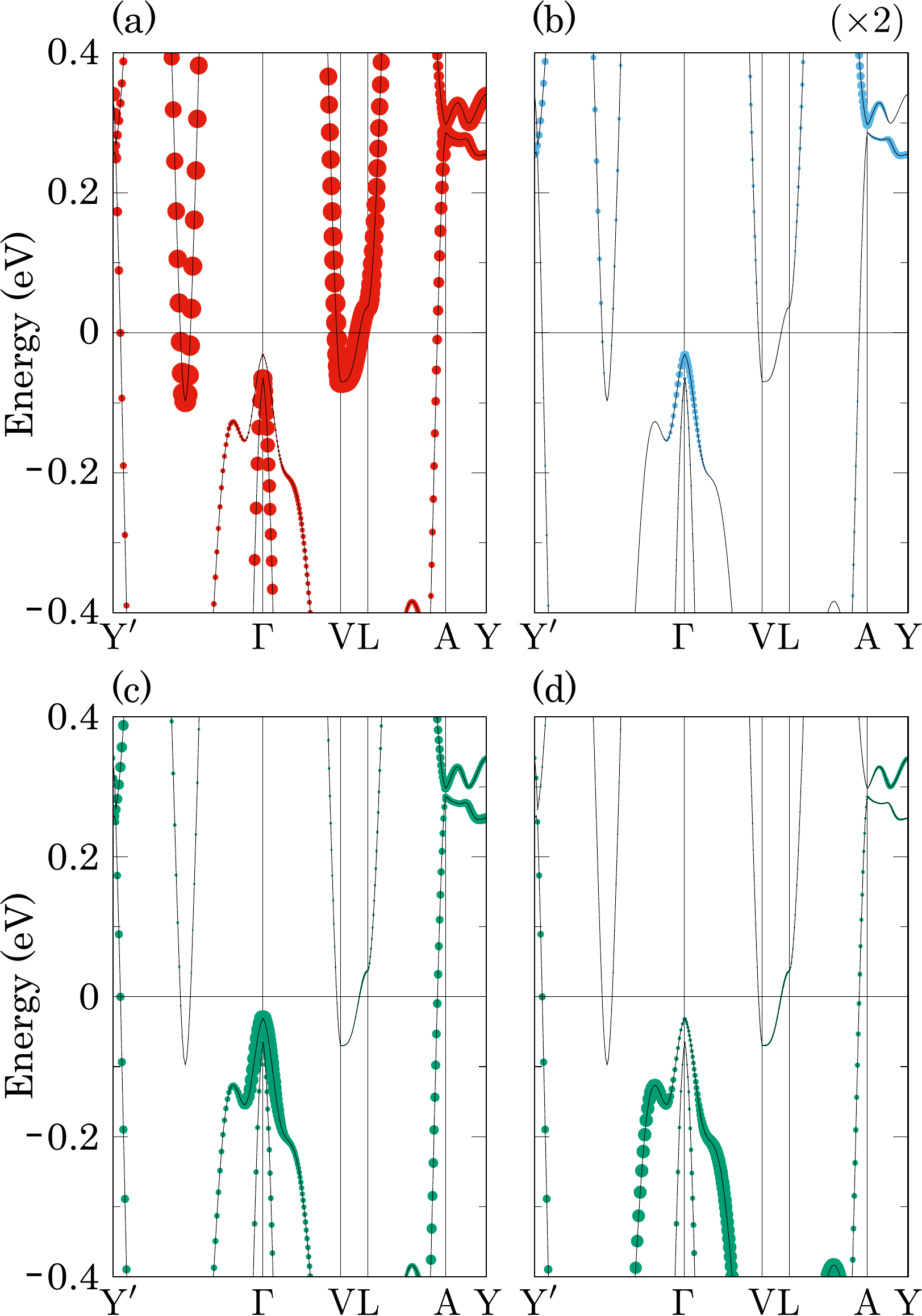}
  \caption{
  Electronic band structure with a weight 
  of (a) Ta-$d$, (b) Pd-$d$, (c) Se(1)(2)-$p$, and (d) Se(3)-$p$ orbitals.
  These band dispersions were calculated with the tight-binding model consisting of the Wannier orbitals.
  For panel (b), the orbital weight is enlarged with a factor of two for clarity.
  }
  \label{fig:band_orbital}
  \end{center}
\end{figure}

For calculations of electron-phonon coupling, we constructed Wannier functions as described in Sec.~\ref{sec:methods}. In Figs.~\ref{fig:band}(a)(b), band dispersion obtained by Wannier interpolation is shown with red solid lines, which satisfactorily reproduce the original first-principles band dispersion shown with black broken lines.
Using the tight-binding model consisting of the Wannier functions, we present the electronic band structure with orbital weight in Fig.~\ref{fig:band_orbital}.
The electron pocket around the midpoint of the Y$'$-$\Gamma$ line or around the V-L line, both of which have the same value of $k_y$, mostly consists of Ta-$d$ orbitals.
The hole pocket around the A or Y($'$) point has Ta-$d$ and Se-$p$ weights. The valence bands at the $\Gamma$ point are also characteristic. The highest valence band at the $\Gamma$ point consists mainly of Se(1)(2)-$p$ orbitals with slight hybridization with Pd-$d$ orbitals, which means that this electronic structure originates from PdSe$_4$ chains. On the other hand, the second highest valence band at the $\Gamma$ point mainly consists of Ta-$d$ orbitals.
The highest valence band at the $\Gamma$ point is slightly below the Fermi energy, which is consistent with the recent ARPES study~\cite{Ootsuki2025}.
Although the highest valence band at the $\Gamma$ point does not intersect the Fermi level, it dramatically affects electron-phonon scattering of the electron pocket, as we shall see later in this paper.

\subsection{Electron-phonon scattering\label{sec:self}}

\begin{figure}
  \begin{center}
  \includegraphics[width = 0.9\linewidth]{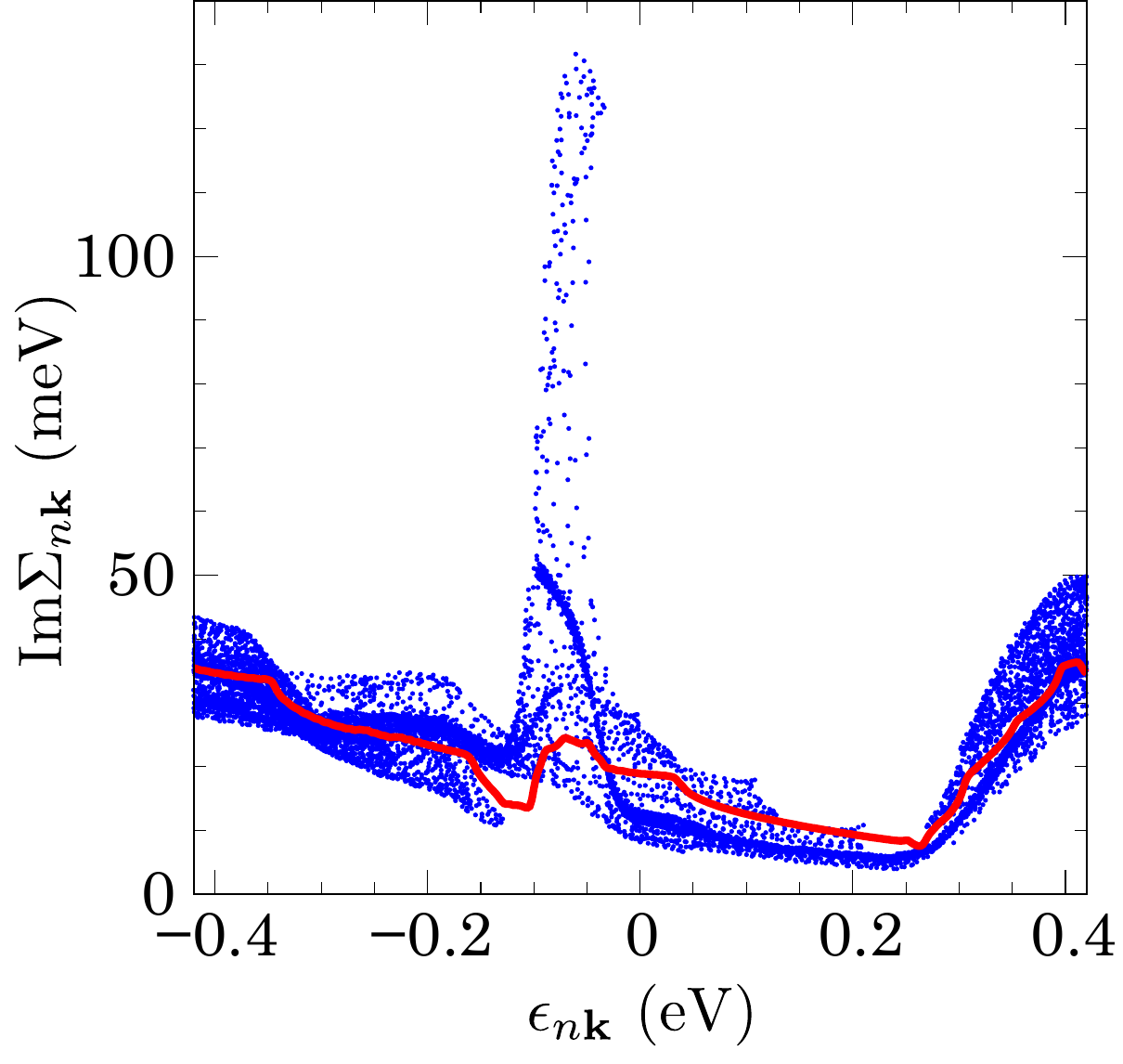}
  \caption{
   The imaginary part of the electron self-energy $\Sigma_{n{\bm k}}$ plotted against the Kohn--Sham energy $\epsilon_{n{\bm k}}$ on an $80\times 80\times 8$ ${\bm k}$ mesh at 300 K. The Fermi level at 300 K was set to zero in the horizontal axis.
   Blue dots represent the calculated values of $\mathrm{Im}\Sigma_{n{\bm k}}$. 
   For comparison, a red line that is proportional to the electronic density of states $\mathrm{DOS}(\epsilon)$ is also shown.
  }
  \label{fig:self_on_kmesh}
  \end{center}
\end{figure}

We calculated the electron self-energy originating from electron-phonon coupling on an $80\times 80\times 8$ ${\bm k}$ mesh at 300 K.
The imaginary part of the calculated self-energy, $\mathrm{Im}\Sigma_{n{\bm k}}$, plotted against the Kohn-Sham energy $\epsilon_{n{\bm k}}$ is shown in Fig.~\ref{fig:self_on_kmesh}.
We also show a red line that is proportional to the electronic density of states $\mathrm{DOS}(\epsilon)$ for comparison.
It is often the case that $\mathrm{Im}\Sigma_{n{\bm k}}$ is roughly proportional to the electronic density of states~\cite{Bernardi2014,Ponce2016,Konstantinou2019,Xia2019_CoSi,Xia2019,Yuan2018,Ha2019} because the imaginary part of the self-energy, Eq.~(\ref{eq:self}), comes down to the Fermi's golden rule with the Bose-Einstein distribution functions as coefficients if the temperature is sufficiently high so that the Bose-Einstein distribution function is dominant in the numerator of the equation.
In that case, if the electron-phonon coupling can be regarded as constant and the electronic states are slowly varying within the phonon energy scale, the imaginary part of the self-energy is roughly proportional to the electronic density of states.
In fact, $\mathrm{Im}\Sigma_{n{\bm k}}$ is roughly approximated by the red line as shown in Fig.~\ref{fig:self_on_kmesh}.
However, remarkably large $\mathrm{Im}\Sigma_{n{\bm k}}$ are found near the Fermi level.
This feature is a hallmark of the strong wave-vector dependence of electron-phonon scattering, and thus we next see the electron self-energy in detail along some ${\bm k}$-path.

\begin{figure}
  \begin{center}
  \includegraphics[width = 0.92\linewidth]{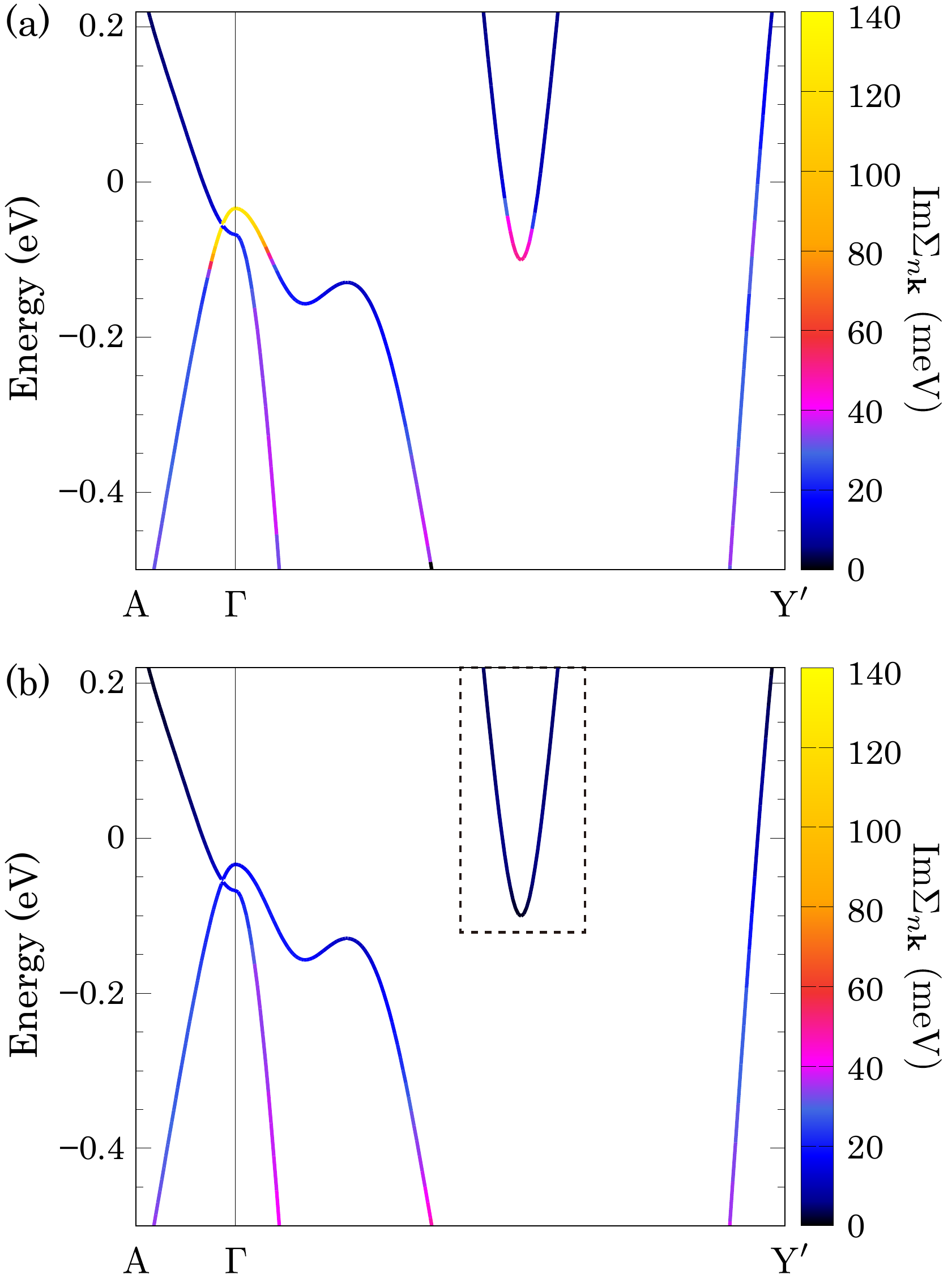}
  \caption{
(a) $\mathrm{Im}\Sigma_{n{\bm k}}$ shown on the electronic band structure at 300 K.
(b) The same as (a), but calculated with the electron pocket shifted upward by $+1$ eV. The inset shown with dotted lines presents the electron pocket that was shifted upward in the calculation of the self energy but shown in the original position in the inset.
For both panels, the Fermi level at 300 K was set to zero in the vertical axis.
  }
  \label{fig:tau_on_band}
  \end{center}
\end{figure}

Figure~\ref{fig:tau_on_band}(a) presents $\mathrm{Im}\Sigma_{n{\bm k}}$ shown on the electronic band structure at 300 K.
It is clearly shown that the highest valence band at the $\Gamma$ point and the bottom of the electron pocket along the $\Gamma$-Y$'$ line suffer from strong scattering.
We also performed the self-energy calculation with an artificial energy shift of $+1$ eV for the electron pocket using the scissor operator~\cite{note_scissor}, as shown in Fig.~\ref{fig:tau_on_band}(b).
Namely, the intervalley scattering between the electron pocket and other bands is artificially excluded in this calculation.
The inset in Fig.~\ref{fig:tau_on_band}(b) shows the electron pocket at the original energy position, while it was shifted upward by $+1$ eV in the self-energy calculation.
In Fig.~\ref{fig:tau_on_band}(b), the highest valence band at the $\Gamma$ point and the electron pocket do not have a large $\mathrm{Im}\Sigma_{n{\bm k}}$, suggesting that the strong scattering discussed in Figs.~\ref{fig:self_on_kmesh} and \ref{fig:tau_on_band}(a) is caused by the intervalley scattering between the highest valence band at the $\Gamma$ point and the electron pocket.

\begin{figure}
  \begin{center}
  \includegraphics[width = 0.5\linewidth]{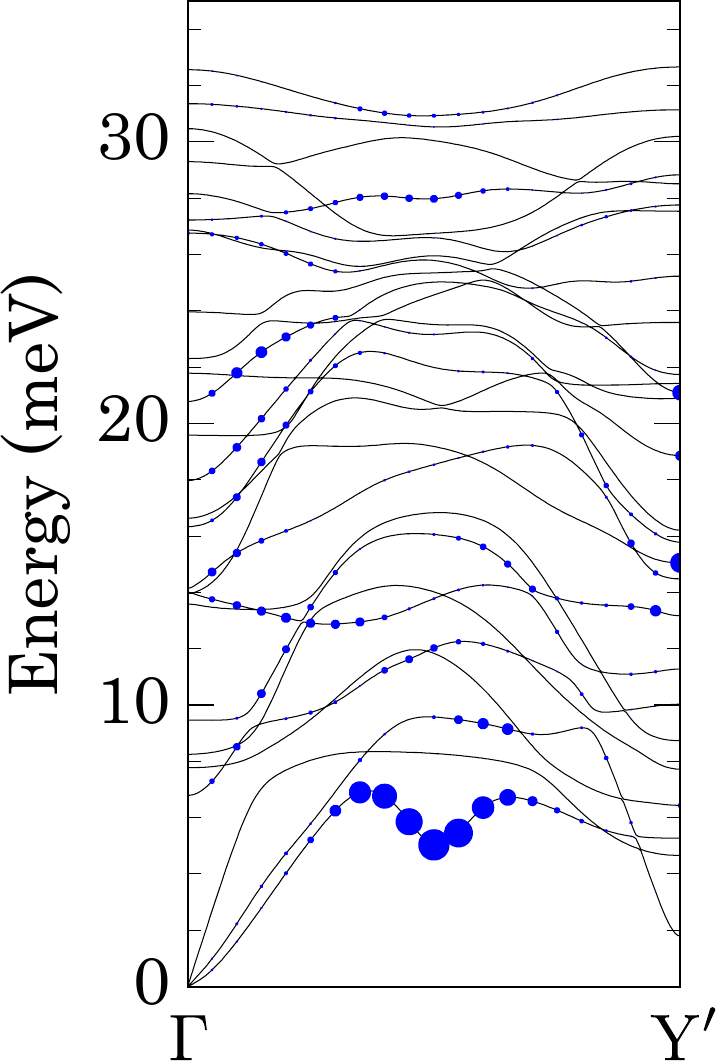}
  \caption{
Absolute value of the electron-phonon coupling $|g_{mn,\nu}({\bm k}, {\bm q})|$ for the specific electronic state indices $n, m, {\bm k}$ plotted as a function of the phonon state indices $(\nu, {\bm q})$ on the phonon dispersion, where the size of blue circles represent the size of $|g|$.
The indices of $g$ are specified as follows:
$n = n_{\mathrm{val}}$, $m=n_{\mathrm{val}}+1$, and ${\bm k}=(0,0,0)$ (see the main text for details).
  }
  \label{fig:g}
  \end{center}
\end{figure}

To investigate the origin of the strong intervalley scattering, we show the electron-phonon coupling $g_{mn,\nu}({\bm k}, {\bm q})$ for some selected states in Fig.~\ref{fig:g}. Since $g$ has many indices, we specified the electronic state indices $n, m, {\bm k}$ and then plotted it as a function of the phonon state indices $(\nu, {\bm q})$ on the phonon dispersion.
The indices are specified as follows: $n = n_{\mathrm{val}}$, $m=n_{\mathrm{val}}+1$, and ${\bm k}=(0,0,0)$, where $n_{\mathrm{val}}$ is the number of valence bands.
Thus, Fig.~\ref{fig:g} represents the relevant electron-phonon coupling for the intervalley scattering contribution to the electron self-energy at the highest valence band at the $\Gamma$ point.

As shown in Fig.~\ref{fig:g}, the highest valence band at the $\Gamma$ point is strongly coupled with the soft phonon mode around the midpoint of the $\Gamma$-Y$'$ line.
First, the coincidence of the ${\bm q}$ vector of the soft phonon and the ${\bm k}$ difference between the electron pocket and the $\Gamma$ point should contribute to this enhancement.
In addition to this, as discussed in Sec.~\ref{sec:bands}, this soft phonon mode mainly consists of atomic displacement in PdSe$_4$ chains. Since the Bloch state at the highest valence band at the $\Gamma$ point also consists of the atomic components of PdSe$_4$ chains [see Fig.~\ref{fig:band_orbital}], the large $|g|$ shown in Fig.~\ref{fig:g} might be partially due to such a large overlap.

\begin{figure}
  \begin{center}
  \includegraphics[width = 0.94\linewidth]{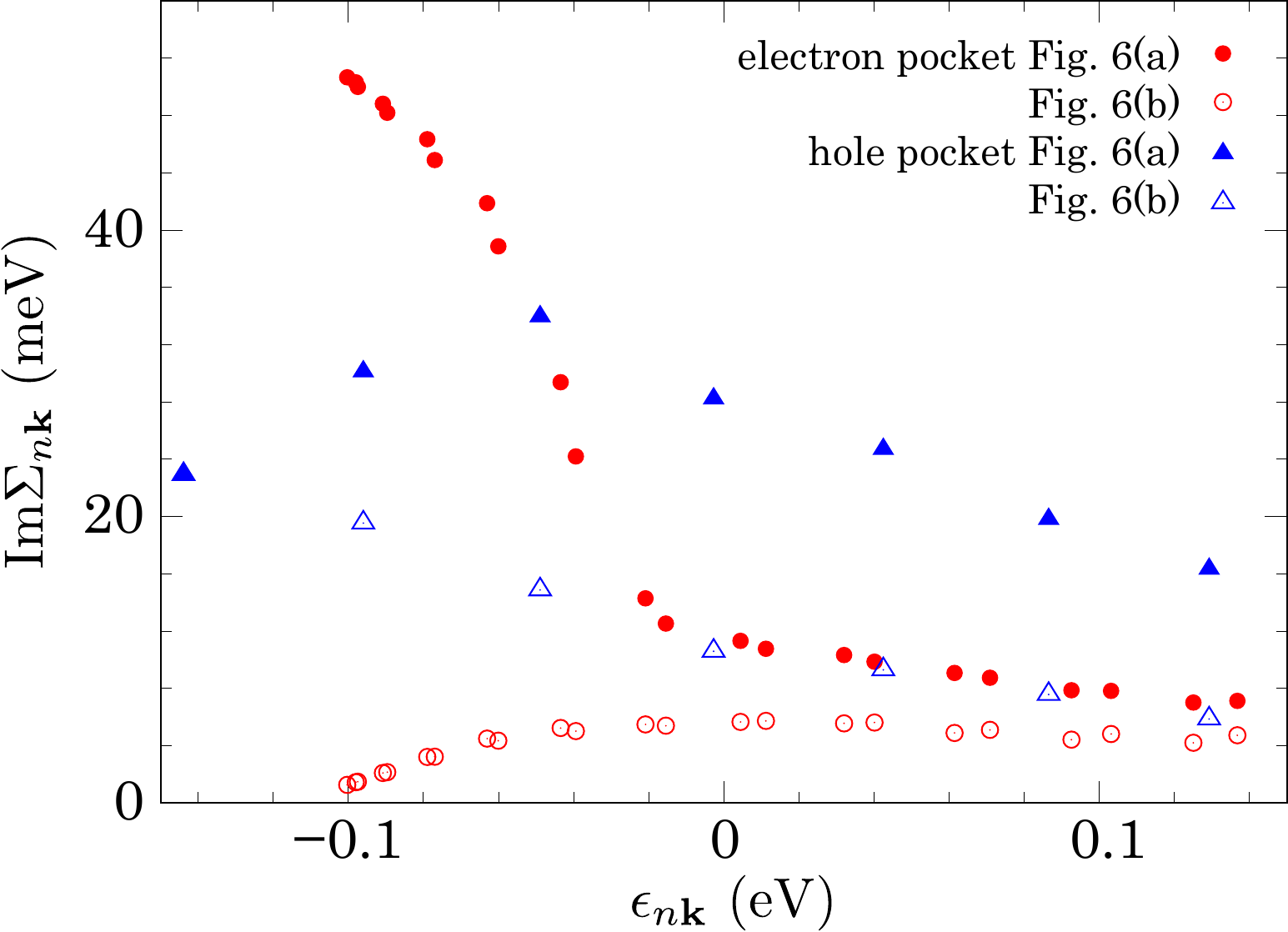}
  \caption{
$\mathrm{Im}\Sigma_{n{\bm k}}$ calculated at 300 K for the Bloch states of the electron (around the midpoint of $\Gamma$-Y) and hole (around Y$'$) pockets along the $\Gamma$-Y$'$ line.
The legends ``Fig.~6(a)'' and ``Fig.~6(b)'' denote electron-phonon coupling calculated with fully considering the electron bands and that calculated without intervalley scattering between electron pockets and valence bands (see the main text and the caption of Fig.~\ref{fig:tau_on_band} for details), respectively.
The Fermi level at 300 K was set to zero in the horizontal axis.
  }
  \label{fig:tau_compare}
  \end{center}
\end{figure}

Since the highest valence band at the $\Gamma$ point lies below the Fermi level and suffers from remarkably strong scattering, Bloch states around the $\Gamma$ point likely contribute less to transport properties.
Thus, in Fig.~\ref{fig:tau_compare}, we show $\mathrm{Im}\Sigma_{n{\bm k}}$ for the Bloch states of the electron and hole pockets, the latter of which is located near the Y$'$ point, along the $\Gamma$-Y$'$ line.
Note that the band dispersion along the $\Gamma$-A line that also crosses the Fermi level and forms the hole pocket, is connected to the hole pocket around the Y point [see, e.g., Ref.~\cite{Nakano2021} for the Fermi surface].

Figure~\ref{fig:tau_compare} shows a comparison of the self-energy between Figs.~\ref{fig:tau_on_band}(a) and \ref{fig:tau_on_band}(b), which allows one to identify the effect of intervalley scattering associated with the electron pocket.
For $\mathrm{Im}\Sigma_{n{\bm k}}$ for carriers in the electron pocket, the effect of the intervalley scattering is significant at $\epsilon_{n\bm{k}}< 20$ meV, which coincides with the energy region where the highest valence band resides at the $\Gamma$ point. This observation again suggests that intervalley scattering between the electron pocket and the $\Gamma$ point is crucial.
We can see that hole carriers also suffer from the intervalley scattering with the electron pocket, while the energy dependence of $\mathrm{Im}\Sigma_{n{\bm k}}$ is moderate compared with that for the electron pocket.

\begin{figure}
  \begin{center}
  \includegraphics[width = 0.84\linewidth]{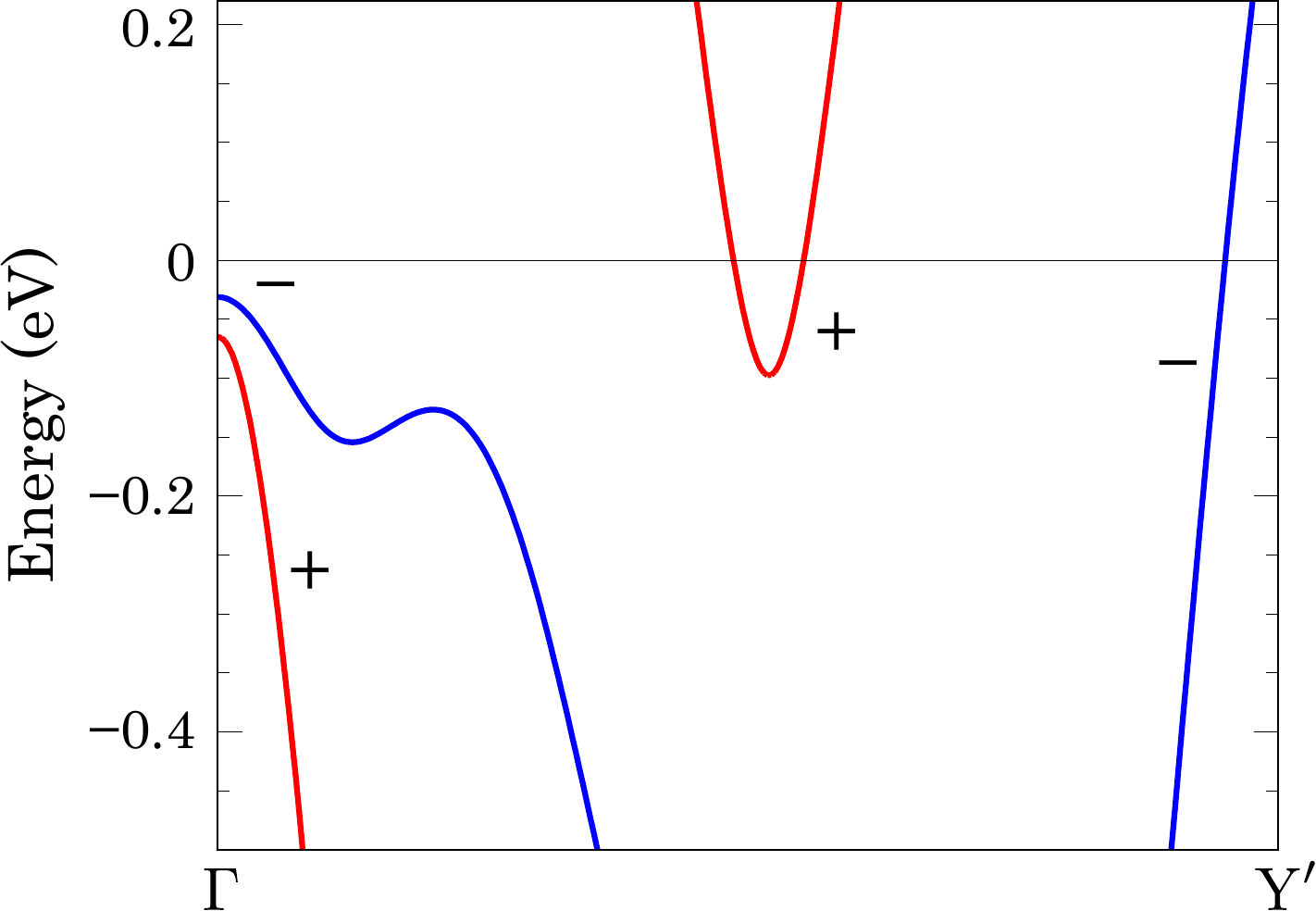}
  \caption{
Electronic band structure along the $\Gamma$-Y$'$ line colored with the eigenvalue with respect to the $C_2$ rotational symmetry in the $xz$ plane, $(x,y,z) \to (-x, y,-z)$. Red and blue denote states with $+1$ or $-1$ eigenvalues, respectively. The Fermi energy was set to zero in the vertical axis.
  }
  \label{fig:band_parity}
  \end{center}
\end{figure}

Finally, we make a short comment on the effect of rotational symmetry on electron-phonon scattering.
In the present system, the $C_2$ rotational symmetry in the $k_xk_z$ plane, $(k_x,k_y,k_z) \to (-k_x, k_y,-k_z)$, belongs to the small group of each ${\bm k}$-point along the $\Gamma$-Y$'$ line.
In other words, this symmetry operation does not change the ${\bm k}$-point along the $\Gamma$-Y$'$ line since $k_x=k_z=0$, and thus Bloch states there can be classified into those with $\pm 1$ eigenvalues with respect to this symmetry [more rigorously, $(x,y,z) \to (-x, y,-z)$]. Figure~\ref{fig:band_parity} illustrates this classification and shows that the highest valence band at the $\Gamma$ point and the electron pocket have different eigenvalues for the rotational symmetry, which activates electron-phonon scattering by the soft phonon with an eigenvalue of $-1$ [see, Sec.~\ref{sec:bands}].
On the other hand, the second highest valence band at the $\Gamma$ point and the electron pocket have the same eigenvalue for the rotational symmetry, by which such scattering is prohibited by symmetry.
This might explain why the second highest valence band at the $\Gamma$ point does not have a large $\mathrm{Im}\Sigma_{n{\bm k}}$ in contrast to the highest valence band in Fig.~\ref{fig:tau_on_band}(a).

Note that this symmetry classification does not make sense for general ${\bm k}$ points that are not preserved against the $C_2$ rotational symmetry. Nevertheless, due to quasi-one-dimensionality along the chain ($k_y$) direction (also see, e.g., Ref.~\cite{Nakano2021} for the quasi-one-dimensional Fermi surface), this kind of suppression is expected to work to some extent in the Brillouin zone or at least for some small portion of the pockets.

\section{Discussion\label{sec:discussion}}

In this paper, we have investigated the valley-selective electron-phonon scattering in Ta$_2$PdSe$_6$. Although transport properties such as the electrical conductivity and the Seebeck coefficient have received a great deal of attention in experiments, transport calculations to obtain these quantities require a very fine ${\bm k}$-mesh especially at low temperatures.
Due to the expensive computational cost for transport calculations considering electron-phonon scattering in the present system, we regard such calculations as an important future issue.
Nevertheless, here we mention how our calculation results shown in the present paper can be related to transport properties, since we believe that our calculation results offer useful information for understanding the transport properties of this material.

In Fig.~\ref{fig:tau_compare}, we have shown that $\mathrm{Im}\Sigma_{n{\bm k}}$ exhibits a sharp drop near the Fermi level for the electron pocket while that for the hole pocket has rather moderate energy dependence.
It has been pointed out that such a sharp drop in carrier lifetime near the Fermi level can effectively mask carriers below the Fermi level and thus enhance the Seebeck effect, as described in the introduction.
In other words, energy filtering can take place for the electron pocket, and thus the negative Seebeck coefficient will be enhanced when electron-phonon scattering is sufficiently active, i.e., at high temperatures. In fact, experimental studies~\cite{Nakano2021, Nakano2021_JPSJ, Nakano2022} show that the Seebeck coefficient becomes negative for $T>100$ K and reaches $\sim -30\ \mu$V/K at room temperature for (Ta$_{1-x}$Nb$_x$)$_2$PdSe$_6$~\cite{Nakano2022}. The enhancement of the negative Seebeck coefficient can partially due to the energy filtering effect of the electron pocket owing to the strong intervalley scattering with the $\Gamma$-point valence band via electron-phonon coupling with the soft phonon.
The enhancement of the Seebeck coefficient due to the strongly energy-dependent scattering has been discussed in much literature~\cite{Xia2019_CoSi,Xia2019,Kumarasinghe2019,Fedorova2022,Kumarasinghe2019,Fedorova2022,Ochi2023,Graziosi2024,Garmroudi2023, Riss2024,Garmroudi2025,Garmroudi2025_Ge,Tohyama2025}. Electron-hole asymmetry is important also in these cases, while the term ``electron (hole) carriers'' does not denote carriers in electron (hole) pockets but means carriers above (below) the Fermi level.

On the other hand, the positive Seebeck coefficient at very low temperatures, which reaches 40 $\mu$V/K at 20 K~\cite{Nakano2021}, has drawn much attention in experiments.
In our calculation, the soft phonon has an energy of 5 meV, which would become inactive at such a low temperature.
In that case, carriers in the electron and hole pockets would have a similar lifetime, by which the Seebeck coefficient would become small.
Is there a possible way to reproduce the short lifetime of electron carriers even at low temperatures, in contrast to the much longer lifetime of hole carriers naturally expected for electron-phonon scattering, as observed in experiments~\cite{Nakano2022,Nakano2025}?
Interestingly, recent inelastic X-ray scattering experiments find that the soft phonon mode exhibits strong anharmonicity and enhanced softening at low temperatures~\cite{Nakano_xray}.
Since our calculation shows that electron pockets have strong coupling with the soft phonon mode, it perhaps be the case that electron carriers exhibit some anomalously enhanced scattering through lattice instability.
Another possibility is that some Bosonic fluctuations that produces a replica band only for the electron pocket as observed by ARPES measurement~\cite{Ootsuki2025} may contribute to electron-carrier scattering.
These possibilities are difficult to address within the present theoretical framework and should be considered as important and challenging problems for future studies.

\section{Summary\label{sec:summary}}

We have investigated electron-phonon scattering in thermoelectric semimetal Ta$_2$PdSe$_6$.
There is a soft phonon mode around $q_y\sim \pi/b$ that mainly consists of atomic displacements in PdSe$_4$ chains.
This soft mode is strongly coupled with the highest valence band at the $\Gamma$ point, which lies slightly below the Fermi energy, and causes strong electron-phonon scattering.
The bottom of the electron pocket energetically overlapped with that band also suffers from strong intervalley scattering, by which the imaginary part of the electron self-energy exhibits a sharp change near the Fermi level.
Although intervalley scattering between the electron and hole pockets is also active, the imaginary part of the self-energy for hole carriers rather shows a moderate energy dependence in contrast to electron carriers.
We have found that electron-phonon scattering is strongly valley-dependent.
Our finding will help us to understand the unique transport properties observed in Ta$_2$PdSe$_6$.

\section*{Acknowledgment}
This study was supported by a research grant from The Thermal and
Electric Energy Technology Foundation and JPSJ KAKENHI Grant No. JP23K13059, No. JP24H01621, No. JP25K08457, and No. JP25K23349.
The computing resource was supported by the supercomputer system in the Institute for Solid State Physics, the University of Tokyo.

\section*{Data Availability}
The data that support the findings of this article are openly available~\cite{data_support}.

\bibliography{main}

\begin{thebibliography}{56}
\bibitem [{not()}]{note_scissor}%
  \BibitemOpen
  \href@noop {} {}\bibinfo {note} {The scissor operator also affects the
  Fermi--Dirac distribution function in Eq.~(\ref{eq:self}). Nevertheless, in
  the room temperature, $T=300$ K, which is much higher than phonon energies,
  the Bose--Einstein distribution function is dominating in the numerator of
  Eq.~(\ref{eq:self}). Thus, the effect of the scissor operators on the
  self-energy, except for the removal of intervalley scatterings between
  valence and conduction bands, is negligible.}\BibitemShut {Stop}%
\end{thebibliography}
\end{document}